\documentclass[preprint,12pt,authoryear]{elsarticle}

\usepackage{amssymb}
\usepackage{amsmath}
\usepackage{bm}
\usepackage{longtable}
\usepackage{comment}
\usepackage{xcolor}

\journal{Icarus}

\begin{document}

\begin{frontmatter}


\author{Kenji Kurosaki\corref{cor1}\fnref{label2}}
\ead{kkurosaki@nda.ac.jp}
\cortext[cor1]{Corresponding author}

\title{Numerical Simulation of Impact Cratering and Induced Seismic Waves in Sand Targets}


\author[label1]{Masahiko Arakawa} 

\affiliation[label2]{organization={Department of Applied Physics, National Defense Academy},
            addressline={Hashirimizu 1-10-20}, 
            city={Yokosuka},
            postcode={239-8686}, 
            state={Kanagawa},
            country={Japan}}
            
\affiliation[label1]{organization={Graduate School of Science, Kobe University},
            addressline={1-1, Rokkodaicho, Nada}, 
            city={Kobe},
            postcode={657-0017}, 
            state={Hyogo},
            country={Japan}}

\begin{abstract}
Impact cratering plays a crucial role in shaping the surfaces of small bodies, satellites, and planets, providing insights into their formation and the history of the Solar System.
Small bodies are often covered with low-cohesion regolith.
Using sand as a model of regolith, we constructed a numerical model for simulating impact on a sand target to investigate the mechanisms of crater formation and impact-induced seismic waves.
Soda-lime glass and quartz sand targets were used for comparison.
The developed sand model successfully reproduced the sound velocity measured in an experimental study.
Using the new sand model, the crater formation was simulated using Smoothed Particle Hydrodynamics with a material strength parameter.
The crater radius and $\pi$-scaling law derived from the numerical simulation were consistent with the experimental study.
The vertical acceleration around the surface of the crater was consistent with the experimentally measured acceleration for the impact-induced seismic wave.
The developed model can provide insight for predicting the size of craters on unknown small bodies.

\end{abstract}




\end{frontmatter}



\section{Introduction}

Impact cratering is an important process that determines the surface condition and features of small bodies, satellites, and terrestrial planets in our Solar System. 
The surface conditions of small bodies, especially the crater shape, provide essential information for understanding how such bodies are formed.
Because small bodies are good analogs of planetesimals, their bulk composition and internal structures indicate the history of the Solar System \cite[e.g.,][]{Watanabe2023}.
Hayabusa 2 and OSIRIS-REx demonstrated that the surface layer of a small body is covered with regolith, where these regolith particles, consisting of grains with high porosity and low cohesion \citep[e.g.,][]{Watanabe2019,Lauretta2019}, were collected.
These small bodies known as rubble pile objects and are thought to have been formed by the reaccumulation of fragments generated by the catastrophic impact \cite[e.g.,][]{Farinella1982}.
Therefore, impact phenomena on low-cohesion granular targets such as regolith are expected to occur during the formation. 

Impacts on small bodies surrounded by regolith material have been the subject of experimental studies in recent years. 
The process of forming an impact crater on the surface of Ryugu was successfully executed and observed via a crater formation experiment using the Small Carry-on Impactor (SCI) on Hayabusa 2 \citep{Arakawa2020}. 
The Double Asteroid Redirection Test (DART) suggests that Dimorphos was covered with a cohesionless regolith material that was ejected sufficiently to change its orbital period around Didymos \citep{Daly2023}.
The shape of Dimorphos was sufficiently changed by DART impact \citep{Raducan2024}.
Because the impact conditions of SCI and DART, such as the mass, velocity, and angle of the impactor, are known, the size of the craters formed by SCI and DART enable identification of the surface material.
During crater formation, the seismic wave induced by the impact propagates around the crater and affects its morphology \citep{Nishiyama2021}.
The impact-induced seismic wave is excited on the surface during crater formation, which is expected to be related to the physical properties of the regolith \citep{McGarr1969,Yasui2015,Matsue2020,Quillen2022,Neiderbach2023,Suo2024}.
\cite{Yasui2015} reported that the velocity and amplitude of impact-induced seismic waves were comparable to the measured speed of sound in the sand material, suggesting that the seismic wave reflects the effective bulk modulus of the sand because the speed of sound propagating in the granular material is about one order of magnitude slower than that in rocks \citep[e.g.,][]{Teramoto2005}.

In a theoretical model based on a hydrodynamic simulation considering the material strength, the sand materials were modeled by a porous rock material incorporated into the equation of state \citep{Wunnemann2008,Jutzi2008,Collins2011b,Raducan2019,Luther2022,Kurosaki2025}.
The discrete element method (DEM) was also used to reproduce the crater experiments, although the impact velocity was less than 1 km s$^{-1}$ \citep{Wada2006,Schwartz2014}.
Using the hydrodynamic model, the effect of porosity was determined by introducing this effect into the equation of state, as in the case of the $P-\alpha$ model, where $P$ is the pressure and $\alpha$ is the distention parameter \citep[e.g.,][]{Herrmann1969,Carroll1972}.
The $P-\alpha$ model requires parameters such as the compaction exponent, full compaction pressure, and pressure and density at the transition between elastic and plastic behavior.
These parameters are determined by comparison with experimental studies \citep{Feldgun2024}.
In contrast, the effect of porosity in the DEM method is determined base on the particle configuration \citep{Makse1999,Velicky2002}.
Analysis of sound transmission in a granular material by numerical simulation showed that the sound velocity depends on $P^{1/6}$ where $P$ is the total pressure, that is, the sum of the confining pressure plus wave-induced pressure \citep{Sanchez2022}.
Consistent with this observation, the sand target should be modeled such that the sound velocity exhibits pressure dependence.
The speed of sound in the sand material was measured for beach sand, demonstrating that the sound speed increases as the depth increases \citep{Bachrach2000}.
DEM simulation suggests that the sound speed is a function of the confined pressure.
The physical properties of sand materials were investigated using models comprising close-packed spheres, and the bulk and shear modulus of the granular material were also predicted \citep{Mogami1965,Oda1974,Digby1981,Walton1987}.
The close-packed sphere models showed good agreement with the sound velocity in unconsolidated sands \citep[e.g.,][]{Bachrach2000}.
Impact phenomena are outside the scope of such packed sphere models of granular materials.
Nevertheless, the models should be valid when the impact velocity is less than or comparable to the speed of sound in a granular material.
Although impact cratering causes localized compression near the impact point, leading to porosity reduction and minor fracturing, the granular model remains applicable for low-velocity impacts, specifically, impacts with subsonic velocities on granular targets, where compressive effects are not dominant.

In this study, a sand model incorporating the granular material model is constructed and the crater formation process is evaluated by comparison of numerical simulations and laboratory experiments, focusing on low-velocity impacts.
Furthermore, the obtained results are compared with the experimental results for impact-induced seismic waves.

The paper is organized as follows:
The numerical model and settings are described in Section \ref{sec:model}.
We present the results of the impact simulation in Section \ref{sec:result} and discuss the implications in Section \ref{sec:discussion}.
Finally, we summarize the findings in Section \ref{sec:conclusion}.

\section{Method} \label{sec:model}
\subsection{SPH method}
To investigate the crater formation, we use the smoothed particle hydrodynamics for elastic dynamics \citep{Libersky1991,Benz1994,Monaghan2000}, hereafter elastic SPH. We use the SPH method developed by \citet{Sugiura2018}. 

Below, we summarize the SPH simulation.
The equations for an elastic body are as follows:
\begin{eqnarray}
    \frac{d\rho}{dt} &=& -\rho\frac{\partial v^\alpha}{\partial x^\alpha} \label{EOC} \\
    \frac{d\bm{v}^\alpha}{dt} &=& \frac{1}{\rho}\frac{\partial \sigma^{\alpha\beta}}{\partial x^\beta} + g^{\alpha} \delta^{\alpha z} \label{EOM} \\
    \frac{du}{dt} &=& \frac{1}{\rho}\sigma^{\alpha\beta}\frac{\partial v^\alpha}{\partial x^\beta} \label{EOE}
\end{eqnarray}
where the stress tensor is represented as:
\begin{equation}
    \sigma^{\alpha\beta} = -p \delta^{\alpha\beta} + S^{\alpha\beta}. \label{stress}
\end{equation}
$t$ is the time, $\rho$ is the density, $\bm{v}$ is the velocity vector, $\bm{x}$ is the position vector, $\sigma^{\alpha\beta}$ is the stress tensor, $u$ is the specific internal energy, $p$ is the pressure, $S^{\alpha\beta}$ is the deviatoric stress tensor, $g^\alpha$ is the gravity for $\alpha$ component, and $\delta^{\alpha\beta}$ is the Kronecker delta. 
Superscripts indicate directions or components of a vector or tensor, where $\alpha, \beta, \gamma=x,y,z$. Equation \ref{EOM} accounts for gravitational effects in the z-direction.
The summation rule over repeated indices in superscripts, indicated by Greek letters, is applied. 
To treat elastic bodies using the SPH method, we use the following equations:
\begin{eqnarray}
    \frac{\mathrm{d}\rho_i}{\mathrm{d}t} &=& -\sum_j m_j\frac{\rho_i}{\rho_j} \left( v_j^\alpha - v_i^\alpha \right) \frac{\partial}{\partial x_i^\alpha} W(|\bm{x}_i-\bm{x}_j|,h), \label{eq_eoc} \\
    \frac{\mathrm{d}v_i^\alpha}{\mathrm{d}t} &=& \sum_j m_j \left[ \frac{\sigma_i^{\alpha\beta}}{\rho_i^2} + \frac{\sigma_j^{\alpha\beta}}{\rho_j^2} -\Pi_{ij}\delta^{\alpha\beta} \right] \frac{\partial}{\partial x_i^\beta} W(|\bm{x}_i-\bm{x}_j|,h) + g^\alpha \delta^{\alpha z}, \label{eq_eom}   \\
    \frac{\mathrm{d}u_i^\alpha}{\mathrm{d}t} &=& -\sum_j \frac{m_j}{2} \left[ \frac{p_i}{\rho_i^2} + \frac{p_j}{\rho_j^2} +\Pi_{ij}\delta^{\alpha\beta} \right] \left( v_j^\alpha - v_i^\alpha \right) \frac{\partial}{\partial x_i^\beta} W(|\bm{x}_i-\bm{x}_j|,h) \\
    && +\sum_j \frac{m_j}{2} \frac{S_i^{\alpha\beta}}{\rho_i \rho_j} \left[\left( v_j^\alpha - v_i^\alpha \right) \frac{\partial}{\partial x_i^\beta} + \left( v_j^\beta - v_i^\beta \right) \frac{\partial}{\partial x_i^\alpha}  \right] W(|\bm{x}_i-\bm{x}_j|,h). \label{eq_eoe}
\end{eqnarray}
where subscripts are the particle number, $m_i$ is the mass of the $i$th SPH particle, 
${v}_i^\alpha$ is its velocity vector for the $\alpha$ component, $x_i^\alpha$ is its position vector for the $\alpha$ component, $h$ is thesmoothing length, 
$W(|\bm{x}_i-\bm{x}_j|,h)$ is a kernel function, and $\Pi_{ij}$ is the artificial viscosity.
The kernel function adopts a Gaussian kernel given by 
\begin{equation}
    W(|\bm{x}_i-\bm{x}_j|,h) = \left[ \frac{1}{h\sqrt{\pi}} \right]^3 \exp\left( -\frac{(x_i^\alpha-x_j^\alpha)^2}{h^2} \right). \label{kernel}
\end{equation}
We set the smoothing length to the initial average particle spacing, which we set to a constant value because the density does not change significantly.
The artificial viscosity works only if the $i$th and $j$th particles are approaching each other, that is $(\bm{v}_i-\bm{v}_j)\cdot(\bm{x}_i-\bm{x}_j)<0$.
In this case, the artificial viscosity is given by:
\begin{eqnarray}
    \Pi_{ij} &=& \frac{-\alpha_\mathrm{vis}\mu_{ij}(C_{s,i}+C_{s,j})/2 + \beta_\mathrm{vis}\mu_{ij}^2}{(\rho_i+\rho_j)/2}, \label{artifical_vis} \\
    \mu_{ij} &=& \frac{h(v_i^\alpha -v_j^\alpha)\cdot(x_i^\alpha-x_j^\alpha)}{(x_i^\alpha-x_j^\alpha)^2 + 0.01h^2}, \label{artificial_vis_mu}
\end{eqnarray}
otherwise $\Pi_{ij} = 0$.
Here, $\alpha_\mathrm{vis}$ and $\beta_\mathrm{vis}$ are parameters for the artificial viscosity. We adopt $\alpha_\mathrm{vis}=1.0$ and $\beta_\mathrm{vis}=2.0$.
The deviatoric stress is proportional to the strain, according to Hooke’s law.
To calculate the deviatoric stress, the time evolution of the strain is obtained after which the time evolution of the deviatoric stress is determined as:
\begin{equation}
    \frac{\mathrm{d}S_i^{\alpha\beta}}{\mathrm{d}t} = 2G\left( \varepsilon_i^{\alpha\beta} -\frac{1}{3}\varepsilon_i^{\gamma\gamma}\delta^{\alpha\beta} \right) + S_i^{\alpha\gamma} R_i^{\beta\gamma} + S_i^{\beta\gamma} R_i^{\alpha\gamma}, \label{hooklaw}
\end{equation}
where $G$ is the shear modulus, $\varepsilon_i^{\alpha\beta}$ and $R_i^{\alpha\beta}$ are the strain rate tensor and rotational rate tensor, respectively, and are represented as:
\begin{eqnarray}
    \varepsilon_i^{\alpha\beta} &=& \frac{1}{2} \left( \frac{\partial v_i^\alpha}{\partial x_i^\beta} + \frac{\partial v_i^\beta}{\partial x_i^\alpha} \right), \label{eq_eps} \\
    R_i^{\alpha\beta} &=& \frac{1}{2} \left( \frac{\partial v_i^\alpha}{\partial x_i^\beta} - \frac{\partial v_i^\beta}{\partial x_i^\alpha} \right). \label{eq_rotational}
\end{eqnarray}
We parallelized our simulation code using the Framework for Developing Particle Simulator \citep[FDPS;][]{Iwasawa2016,Namekata2018}.

\subsection{Friction model}
To represent the friction of granular materials, we used the Drucker-Prager model to determine the yielding strength $Y_i$ represented as
\begin{equation}
    Y_i = \min\left( Y_0 + f p_i, Y_\mathrm{max} \right), \label{DPmodel}
\end{equation}
where $Y_0$ is the cohesion, $f$ is the friction coefficient, and $Y_\mathrm{max}$ is the maximum yielding stress.
In the calculation, we assumed that $Y_0=0$ for whole simulations. The cohesion is very low for the dry sand material but is non-negligible for wet sand \citep[e.g.,][]{LaJeunesse2017}.
To use the yielding strength from Eq.~\ref{DPmodel}, we modify the deviatoric stress tensor $S_i^{\alpha\beta}$ as:
\begin{eqnarray}
    S_i^{\alpha\beta} &\to& f_i S_i^{\alpha\beta}, \\
    f_i & = & \min\left[ Y_i/\sqrt{J_{2,i}}, 1 \right], \\
    J_{2,i} &=& \frac{1}{2}S_i^{\alpha\beta}S_i^{\alpha\beta}.
\end{eqnarray}
We adopt $f$ by using the measured angle of response $\theta_m$ and
we adopt $f$ by \citep[e.g.,][]{Luther2022} represented as
\begin{equation}
    f = \frac{2\sqrt{2}\sin\theta_m}{3-\sin\theta_m}. \label{Luthertheta}
\end{equation}

The projectile material was modeled using the von Mises yielding criterion \citep[e.g.,][]{Benz1994}, and its mechanical properties were based on the Makrolon\textregistered{} 2407 datasheet, as summarized in Table 2. The same projectile configuration was used for all impact simulations in this study.

\begin{longtable}{cccc}
    \caption{Material properties of granular particles. \label{tab:eosparam}}
    \\
    \hline
         Parameter & Glass & Quartz  \\
    \hline
    \endfirsthead
    \caption[]{Continued from the previous page} \\
    \hline
         Parameter & Soda-lime glass & Quartz sand  \\
    \hline
    \endhead
    \hline    
         Density [g cm$^{-3}$] & 2.5 & 2.2 \\
         Shear modulus [GPa] & 25 & 31 \\
         Poisson ratio & 0.20 & 0.17 \\
         Bulk modulus [GPa] & 33 & 36.6 \\ 
         Friction coefficient & 0.42 & 0.63 \\
        \hline
\end{longtable}

\begin{longtable}{cccc} 
    \caption{Material properties of projectile materials (polycarbonate data are adopted Makrolon\textregistered{} 2407 from \citet{CovestroMakrolon2407})} \label{tab:proj}
    \\
    \hline
         Parameter & Polycarbonate   \\
    \hline
    \endfirsthead
    \caption[]{Continued from the previous page} \\
    \hline
         Parameter & Polycarbonate   \\
    \hline
    \endhead
    \hline    
         Density [g cm$^{-3}$] & 1.2   \\
         Shear modulus [GPa] & 0.93   \\
         Poisson ratio & 0.35  \\
         Bulk modulus [GPa] & 0.95  \\
         Yielding Strength [MPa] & 66 \\
        \hline
\end{longtable}

\subsection{Material model of granular target}
The sand target model is described in this subsection. The assumptions are summarized as follows:
\begin{itemize}
    \item The granular target is composed of closely packed spheres of a constant size. 
    \item The density of the granular target is determined by its porosity.
    \item The spheres comprising the granular target do not undergo disruption.
    \item The spheres are perfectly smooth: thus, the tangential forces among them are ignored.
\end{itemize}

Here, we present the density model used in our simulation. The density of the granular target is represented in terms of the porosity of the granular material.
The intrinsic porosity of the individual particles comprising the granular material is considered negligible because the macroscopic porosity can be easily compacted. Therefore, the density of the granular material is given by
\begin{equation}
    \rho = (1-\phi) \rho_s \label{rho_phi}
\end{equation}
where $\phi$ is the porosity and $\rho_s$ is the density of the sphere.
$\rho_0$ is determined by the initial porosity condition $\phi_0$: thus, $\rho_0=(1-\phi_0) \rho_s$.
The material parameters are summarized in Table~\ref{tab:eosparam}.

The bulk modulus of the granular material is determined by the sphere cluster.
\cite{Walton1987} reported the effective elastic moduli for randomly packed identical elastic spheres. 
We adopt the effective bulk modulus $K_\mathrm{eff}$ and shear modulus $G_\mathrm{eff}$ as:
\begin{eqnarray}
    K_\mathrm{eff} & = & \left[\frac{n^2 (1-\phi)^2 G_s^2 P}{18\pi^2 (1-\nu_s)^2} \right]^{1/3} \label{Keff1} \\
    G_\mathrm{eff} & = & \frac{3}{5} \left\{ \left[\frac{n^2 (1-\phi)^2 G_s^2 P}{18\pi^2 (1-\nu_s)^2} \right]^{1/3} + s_t \left[\frac{3}{2}\frac{n^2 (1-\phi)^2 (1-\nu_s) G_s^2 P}{\pi^2 (2-\nu_s)^3} \right]^{1/3} \right\} \label{Geff1}
\end{eqnarray}
where $n$ is the average number of contact points per sphere, $\phi$ is the porosity, $G_s$ is the rigidity of the sphere, $\nu_s$ is the Poisson ratio of the sphere, and $P$ is the confining pressure.
The shear modulus is described as the summation of the normal and shear stiffness of two spheres, which are represented as the first and second terms in Eq.~\ref{Geff1}, respectively.
The second term in Eq.~\ref{Geff1} depends on the friction between the spheres. 
We introduced a parameter reflecting the dependence on the effect of the shear stiffness $s_t$.
In this study, we assumed $s_t=0$.
In the simulation, we set $n=9$, which is a typical value for $\phi=0.4$ \citep{Smith1929,Field1963}.
\footnote{We denoted $\phi$ as the porosity of the granular material, although \cite{Walton1987} denoted $\phi$ as the volume concentration of spheres.}
In this study, we assumed that the confining pressure is equal to the pressure of hydrostatic compression. 
Note $s_t$ represents the shear stiffness parameter.
By combining Eqs. \ref{Keff1} and \ref{Geff1}, the following relationship is obtained:
\begin{equation}
    G_\mathrm{eff} = \frac{3}{5}\left[ 1+\frac{3(1-\nu_s)}{2-\nu_s} s_t \right]K_\mathrm{eff} \label{GKeff}
\end{equation}
When the Poisson ratio of the sand target is described by $\nu_t$, $s_t$ is represented as:
\begin{equation}
    s_t = \frac{1}{2} \frac{2-\nu_s}{1-\nu_s}\frac{1-4\nu_t}{1+\nu_t}. \label{stparam}
\end{equation}
In this study, we assumed the shear stiffness of the sand target; thus, we set $s_t=0$ and the Poisson ratio as $\nu_t=0.25$.
The effective Poisson ratio of the granular material $\nu_\mathrm{eff}=0.25$, is independent of the material property.
The sound velocities of the longitudinal wave $V_p$ and transverse wave $V_s$ are determined by the density of the granular material (Eq.~\ref{rho_phi}), effective bulk modulus $K_{\text{eff}}$ (Eq.~\ref{Keff1}), and effective shear modulus $G_{\text{eff}}$ (Eq.~\ref{Geff1}).
The values of $K_{\text{eff}}$ and $G_{\text{eff}}$ were calculated using the parameters listed in Table~\ref{tab:eosparam}.
The confining pressure is assumed to be the local hydrostatic pressure at the target location.
Using these quantities, $V_p$ and $V_s$ were computed according to the following equations:
\begin{eqnarray}
    V_p & = & \sqrt{\frac{K_\mathrm{eff}+ \frac{4}{3} G_\mathrm{eff}}{\rho}} \label{Vp} \\
    V_s & = & \sqrt{\frac{G_\mathrm{eff}}{\rho}}. \label{Vs}
\end{eqnarray}
We determined the $P$--$\rho$ relation by integrating the bulk modulus relation.
The bulk modulus is determined as:
\begin{equation}
    K_\mathrm{eff} = \rho \frac{dP}{d\rho}. \label{Keff-prho}
\end{equation}
Combining Eqs.\ref{Keff1} and \ref{Keff-prho}, the density-pressure relation is calculated as:
\begin{equation}
    P=\frac{n G_\mathrm{s}\rho_0}{3\sqrt{2}\pi(1-\nu_s)\rho_s} 
    \left[ \left(\frac{\rho}{\rho_0} \right)^{2/3} -1 \right]^{3/2} \label{PRho-sand}
\end{equation}
where $\rho_0$ is the density before compression. 
Regarding the equation of states, we confirm that Eq.~\ref{PRho-sand} was used in the simulations.
In this formulation, the pressure is computed as a function of the density only, and does not depend on the internal energy. The evolution of the internal energy follows the Hugoniot relation.

Compared to the $P$--$\alpha$ model, which includes compaction effects, the granular model developed herein does not account for irreversible compaction of the porous material.
Instead, the model captures the pressure dependence of the speed of sound in a granular material through Hertzian contact mechanics.
This allows the bulk modulus to vary smoothly with confining pressure, making the model particularly suitable for low-velocity impacts below the speed of sound in the granular material, where grain crushing is negligible.

\section{Result} \label{sec:result}
This section presents the crater formation simulations under various conditions. The simulation results are summarized as follows:
First, the sound velocities derived from the developed equation of state are presented in Section~\ref{sub:vs}.
Second, an example of the simulation of crater formation is presented in Section~\ref{sub:crater_example}.
The simulation was set to model a cylindrical polycarbonate projectile with a height and diameter of 1 cm impacting vertically onto a granular target composed of quartz or soda-lime glass at a velocity of 100 m s$^{-1}$.
Finally, the scaling law for crater formation over impact velocities ranging from 100 m s$^{-1}$ to 1 km s$^{-1}$ is discussed in Section~\ref{sub:scale}.

\subsection{Sound velocity} \label{sub:vs}
\begin{figure}
    \centering
    \includegraphics[bb=0 0 480 360, width=0.5\linewidth]{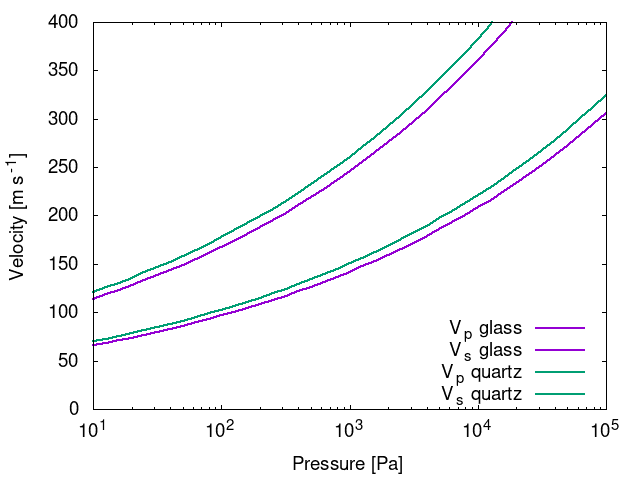}
    \caption{Sound velocity as a function of confining pressure. The colors represent the materials: purple is soda-lime glass beads, and green is quartz sand. The solid and dashed lines are the longitudinal and transverse wave velocities, respectively.}
    \label{fig:cs-sand}
\end{figure}

We assessed the sound velocity using Eqs. \ref{Vp} and \ref{Vs}.
The pressure is equivalent to the depth $z$ calculated by: 
\begin{equation}
    P = \rho g z, \label{pressure_hydro}
\end{equation}
where $g$ is the gravitational acceleration.
In this model, we set the pressure as the sum of the hydrostatic pressure and the wave-induced pressure.
Figure \ref{fig:cs-sand} shows the sound velocity as a function of the total pressure, which is the sum of the confining pressure and the pressure induced by the acoustic wave \citep{Sanchez2022}.
When the granular target is in the hydrostatic condition, the confining pressure should be equal to the hydrostatic pressure. 
The experimentally determined longitudinal wave velocity is $\sim 170$ m s$^{-1}$ \citep[e.g.,][]{Matsue2020}.
This velocity is consistent with the present result within 5\% error for a pressure of 150 Pa and depth equivalent to 1 cm.
On the other hand, the transverse wave velocity estimated using the developed model was overestimated by 50\% compared with the experimental data \citep{Bachrach2000}.
Because constraining the shear stress in the granular target is difficult owing to changes in the contact points, the transverse velocity is difficult to constrain.

\subsection{Crater formation} \label{sub:crater_example}
This subsection describes the simulation of crater formation.
We simulated the impact of a cylindrical polycarbonate projectile on a granular target composed of quartz particles. The initial porosity of the granular target was set to $\phi_0 = 0.4$, and the material parameters for quartz were taken from Table~\ref{tab:eosparam}.
The target was modeled as a hemispherical body with a radius of 15 cm. To impose boundary conditions, particles located in the outer shell region (14.5 to 15 cm from the center) were fixed.
The target was discretized using approximately $10^6$ SPH particles, with a smoothing length of 0.258 cm.
In SPH, the resolution is defined by the particle count and smoothing length, which were chosen to ensure sufficient spatial accuracy. All simulations were performed using double-precision arithmetic to ensure numerical stability and accuracy.
The projectile, comprising polycarbonate (Makrolon\textregistered{} 2407), was modeled as a cylinder with a height and diameter of 1 cm.
The projectile impacted the center of the target in a vertical orientation with an initial velocity of 100 m s$^{-1}$.
The simulation was run for 0.3 s after impact, corresponding to approximately three times the typical formation time for a crater with a diameter of 10 cm.
The crater size was measured based on the rim diameter, defined as the horizontal distance between the highest points on the target surface after impact, relative to the pre-impact surface level.

\begin{figure}
    \centering
    \includegraphics[bb=0 0 2510 1076, width=\linewidth]{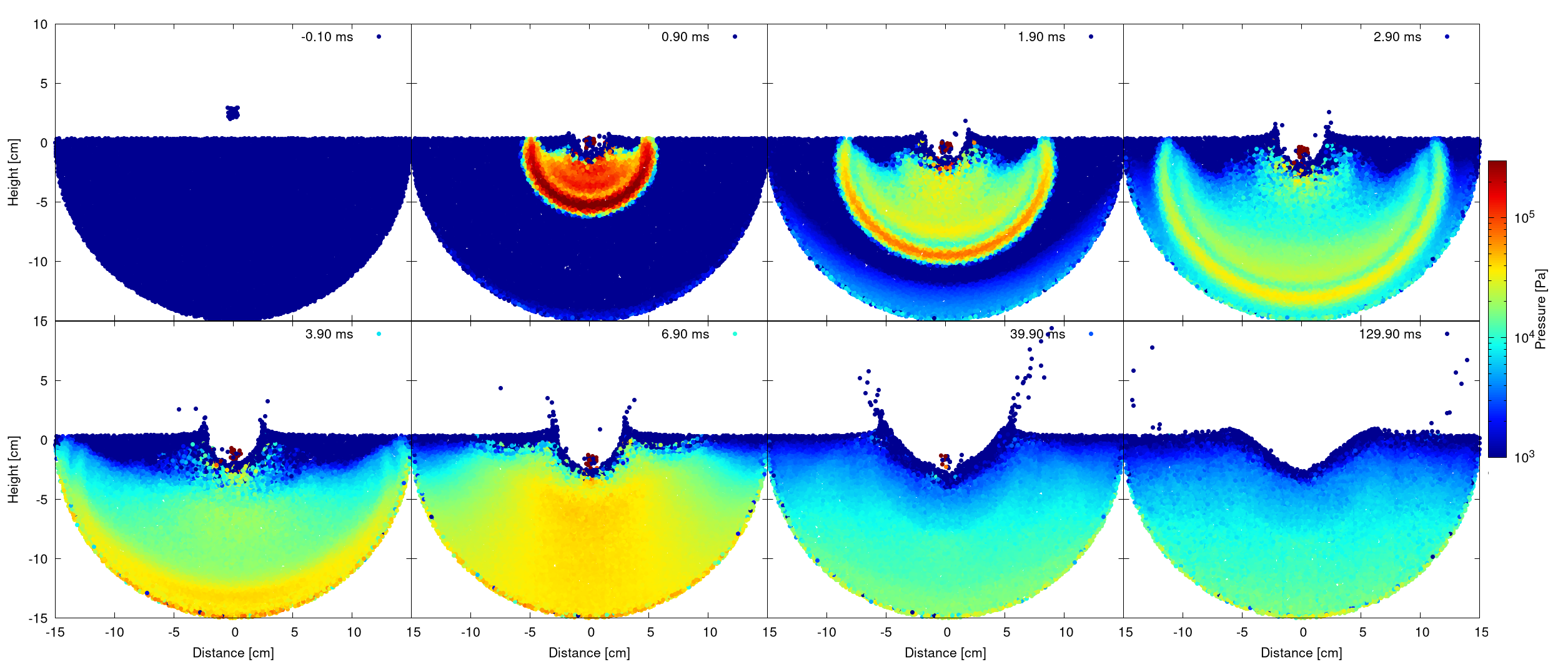}
    \caption{Snapshots of crater formation. This simulation corresponds to RUN 1 in Table~\ref{tab:sims}. The color contour represents the pressure. Dark-blue indicates a the pressure lower than 100 Pa.}
    \label{fig:crater-example}
\end{figure}

\begin{figure}
    \centering
    \includegraphics[bb=0 0 800 600, width=10cm]{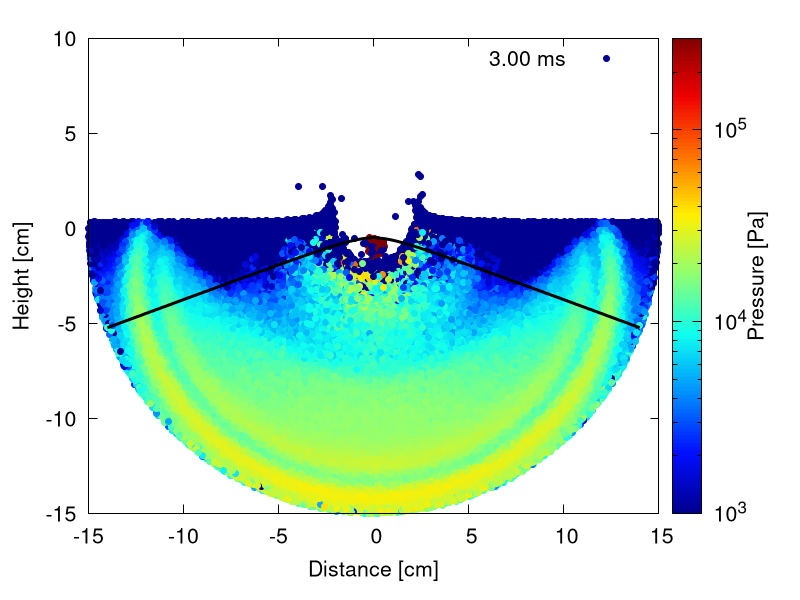}
    \includegraphics[bb=0 0 800 600, width=10cm]{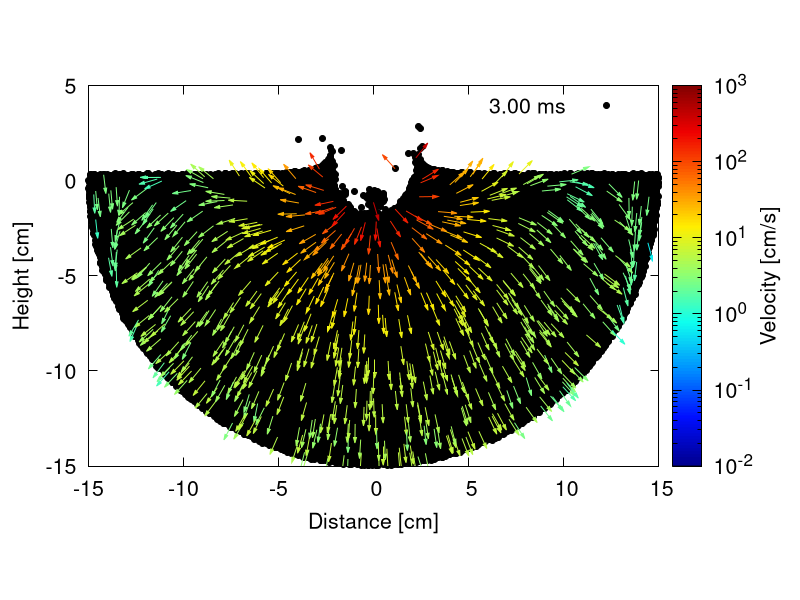}
    \caption{Pressure (upper panel) and velocity (lower panel) distribution for Run 1 at 3 ms. The color bar shows the pressure and absolute value of the velocity, respectively.}
    \label{fig:pandvec}
\end{figure}

\begin{figure}
    \centering
    \includegraphics[bb=0 0 800 600, width=\linewidth]{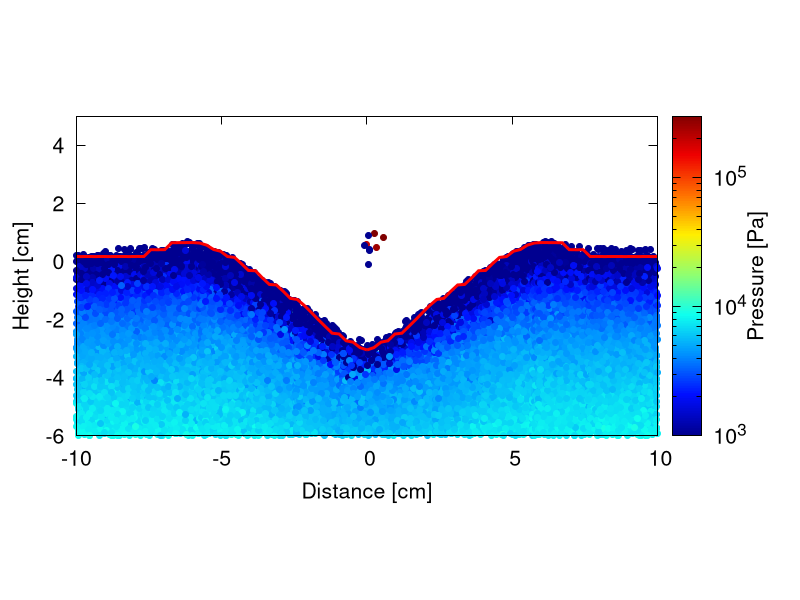}
    \caption{Snapshots of the crater profile for RUN 1. The red line shows the crater profile. The profile shows a cross-section between $y=-0.05$. to $y=0.05$. Particles visible at the center are residual projectile fragments.}
    \label{fig:crater-profile}
\end{figure}

Figure \ref{fig:crater-example} shows the time evolution of the crater formation. 
At $t=2.9$ ms, the wave propagating in the target reached the boundary and reflected to the center of the target.
On the other hand, the wave propagating on the surface reached the boundary at $t = 3.9$ ms.
These differences are caused by irregular shock reflection, which was caused by the peak pressure because the distance from the impact point decreased beyond the irregular shock reflection boundary
\citep[e.g.,][]{Kamegai1986,Kurosawa2018}. 
According to \cite{Melosh1984}, the interference zone $z_p$ is a function of the distance from the impact point $s$, which is described as:
\begin{equation}
    z_p = \frac{c_L\tau}{2} \sqrt{1+\frac{s^2}{d^2-(c_L\tau)^2/2}} \label{zp-def}
\end{equation}
where $d$ is the depth of the equivalent center.
Figure~\ref{fig:pandvec} shows the pressure and velocity distribution at 3 ms.
The surface region is pressure-free; thus, the particles in this region were accelerated and ejected. 
The interference zone provides a good indicator for explaining the pressure-free region until the shock is reflected from the boundary, where spallation occurred when the equivalent depth was assumed to be $d=2 R_p$. 
We also confirmed that the velocity vector near the surface zone tuned upward.
Figure~\ref{fig:crater-profile} shows the crater profile for Run 1. 
The rim-to-rim crater radius and depth were 6 cm and 2.9 cm, respectively, compared with 6.292 cm and 2.6 cm from the experiments. The crater radius was consistent with the experiment, whereas the crater depth was slightly overestimated.
Table~\ref{tab:sims} shows the results of crater formation under the various conditions.
The present simulation reproduced the crater radius observed in experimental studies \citep[see also][]{Yasui2015,Matsue2020}.

\subsection{Scaling law} \label{sub:scale}
\begin{figure}
    \centering
    \includegraphics[bb=0 0 480 360, width=\linewidth]{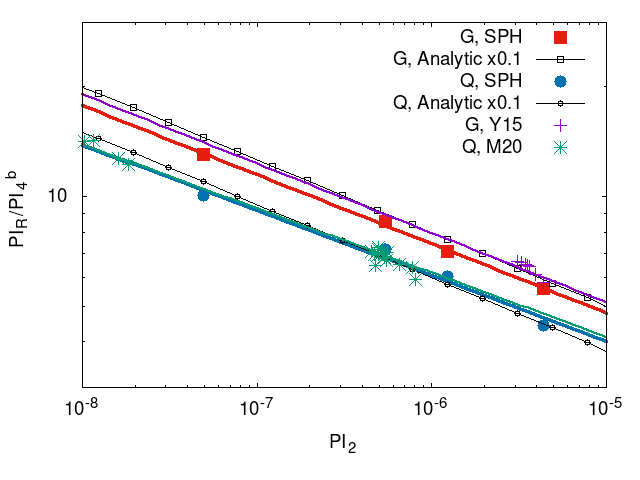}
    \caption{Relationship for $\pi$-scaling law. The red line shows the crater profile. Filled red and blue symbols indicate SPH simulations for soda-lime glass and quartz, respectively. Black lines with open red and blue symbols are the analytical solution calculated using Eq.~\ref{ldcrater} multiplied by 0.1. Purple and green symbols indicate experimental results reported by \cite{Yasui2015} for soda-lime glass (Y15) and \cite{Matsue2020} for quartz (M20), respectively.}
    \label{fig:piscale}
\end{figure}

In this subsection, we derive the $\pi$-scaling law for the glass and quartz targets. 
We used the following relationships between the crater radius, $R$ and the impact conditions described using the non-dimensional scaling parameters $\pi_R$ and $\pi_2$ in the gravity regime:
\begin{eqnarray}
    \pi_R &=& C_0 \pi_2^{a} \pi_4^{b}, \label{piRfit} \\
    \pi_R &=& R\left( \frac{\rho_t}{m_p} \right)^{1/3}, \label{piR} \\
    \pi_2 &=& \frac{gr_p}{v_\mathrm{imp}^2}, \label{pi2} \\
    \pi_4 &=& \frac{\rho_t}{\rho_p}, \label{pi4}
\end{eqnarray}
where $\rho_t$ is the bulk density of the target, $m_p$ is the projectile mass, $g$ is the gravitational acceleration, $C_0$ is a constant, $a = -\frac{\mu}{2+\mu}$, and $b=\frac{2+\mu-6\nu}{3(2+\mu)}$.
Based on the above fitting results, we can determine $\mu$ after calculating the power exponent for $\pi_4$ using $\nu=1/3$ \citep{Housen2003}.
The fitting results are summarized in Table~\ref{tab:fit}.
Figure~\ref{fig:piscale} shows the $\pi$-scaling law for the present numerical simulations versus the experimental results reported by \cite{Yasui2015} and \cite{Matsue2020}.
Although we adopted a simple equation of state for the granular material and a material model, the simulations were in good agreement with the experimental study.
That is, the friction coefficient is an important parameter for determining the crater radius.
We modified the assumptions regarding the initial porosity and Poisson ratio.
The crater radius is influenced by both the initial porosity and the Poisson’s ratio of the granular target \citep[e.g.,][]{Jutzi2008,Raducan2019,Stickle2020}.
Table~\ref{tab:porpoi} presents the crater radius for the granular target composed of glass with various Poisson’s ratios and initial porosities.
Varying the Poisson’s ratio resulted in an approximately 5\% change in the crater radius, whereas changing the initial porosity produced a similar 5\% variation. These effects are minor compared to the influence of the friction coefficient, which remains the dominant factor in determining the crater size.

The applicable range of our findings is limited to subsonic impact velocities, defined here as velocities below the sound speed of the grain. 
This limitation arises because the equation of state used in the model does not account for grain crushing effects. When the local pressure at the impact site exceeds the crushing strength of the particles, the pressure-density relationship changes significantly owing to irreversible compaction and fragmentation.
Experimental studies \citep{Yasui2015,Matsue2020} suggest that grain crushing becomes prominent at impact velocities exceeding approximately 1 km s$^{-1}$ for both soda-lime glass and quartz.
To ensure the validity of the developed model, all simulations were conducted at velocities below this threshold. 

\begin{table}[htbp]
    \centering
    \begin{tabular}{ccccccc}
    \hline
    Run & Material & $v_\mathrm{imp}$ [cm s$^{-1}$] & $R_\mathrm{rim}$ [cm] & $d_c$ [cm] & $\alpha$ & $B$ \\
    \hline
    1 & Glass & 1.0 $\times 10^{4}$ & 6.0 & 2.9  & $-2.46$ & 5.22  \\
    2 &       & 2.0 $\times 10^{4}$ & 6.6 & 3.5  & $-2.51$ & 5.65 \\
    3 &       & 3.0 $\times 10^{4}$ & 9.2 & 3.9  & $-2.54$ & 5.94 \\
    4 &       & 10. $\times 10^{4}$ & 14.0 & 5.9 & $-2.46$ & 6.59   \\
    \hline                         
    5 & Quartz& 1.0 $\times 10^{4}$ & 4.7 & 2.5  & $-2.48$ & 5.19  \\
    6 &       & 2.0 $\times 10^{4}$ & 6.1 & 2.6  & $-2.53$ & 5.61  \\
    7 &       & 3.0 $\times 10^{4}$ & 7.7 & 2.8  & $-2.56$ & 5.89 \\
    8 &       & 10. $\times 10^{4}$ & 10.8 & 4.3 & $-2.57$ & 6.61   \\
    \hline
    \end{tabular}
    \caption{Simulation results of impacts on glass and quartz targets. The run number and the materials are shown. $v_\mathrm{imp}$ is the impact velocity in cm s$^{-1}$, $R_\mathrm{rim}$ is the rim-to-rim crater radius in cm, and $d_c$ is the crater depth from the top of the rim in cm. The target size and smoothing length were 17 cm and 0.258 cm for Run 1--8. The pressure attenuation parameters are also shown. The pressure relation was fitted by $P/P_0 = 10^B (l/r_p)^{-\alpha}$ for each Run when the shock wave had reached the boundary of the target.}
    \label{tab:sims}
\end{table}

\begin{table}[htbp]
    \centering
    \begin{tabular}{ccccc}
    \hline
    Material & $a$ & $b$ & $\mu$ & $C_0$ \\
    \hline
    Glass  & $-0.189$ & 0.0631 & 0.467 & 0.542 \\
    Quartz & $-0.179$ & 0.0596 & 0.436 & 0.512 \\  
    \hline
    \end{tabular}
    \caption{Fitting results with SPH simulation with Eq.~\ref{piRfit}.}
    \label{tab:fit}
\end{table}

\begin{table}[htbp]
    \centering
    \begin{tabular}{ccccc}
    \hline
    Material & Initial porosity $\phi_0$ & Poisson ratio & $R_\mathrm{rim}$ [cm]  \\
    \hline
    Glass  & 0.3 & 0.25 & 5.5  \\
           & 0.5 & 0.25  & 6.4  \\
           & 0.4 & 0.027 & 5.7  \\
            & 0.4  & 0.1 & 6.0  \\  
            & 0.4  & 0.4 & 6.0  \\  
    \hline
    \end{tabular}
    \caption{Crater radius for various initial porosity and Poisson ratio.}
    \label{tab:porpoi}
\end{table}

\section{Discussion} \label{sec:discussion}

\subsection{Impact-induced seismic wave}
In this section, we analyze the impact-induced seismic wave based on comparisons with previous experimental studies. Specifically, we refer to the study of \citet{Yasui2015} in which impacts on granular targets composed of soda-lime glass beads were investigated and that of \citet{Matsue2020} in which similar experiments were conducted using quartz sand.
In both studies, the vertical acceleration was measured using accelerometers placed on the surface of the granular target.
For comparison with these measurements, we extracted the peak vertical and horizontal acceleration values from the present simulations.
We then evaluated the spatial dependence of the peak acceleration values and fitted them to a functional form to characterize the distance-dependent decay of the impact-induced seismic intensity.
For example, in Run 5, we examined the depth dependence of the acceleration signal to assess how the characteristics of seismic waves vary with the accelerometer placement.
These experimental datasets provide benchmark values for validating the simulation results under low-velocity impact conditions.
In the simulations, the projectile was modeled as polycarbonate, whereas different impactor materials were used in the experimental studies depending on the test parameters.

\begin{figure}
    \centering
    \includegraphics[bb=0 0 480 360, width=10cm]{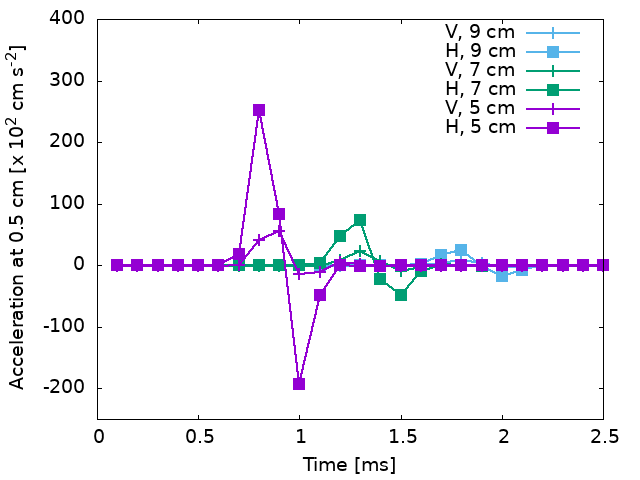}
    \includegraphics[bb=0 0 480 360, width=10cm]{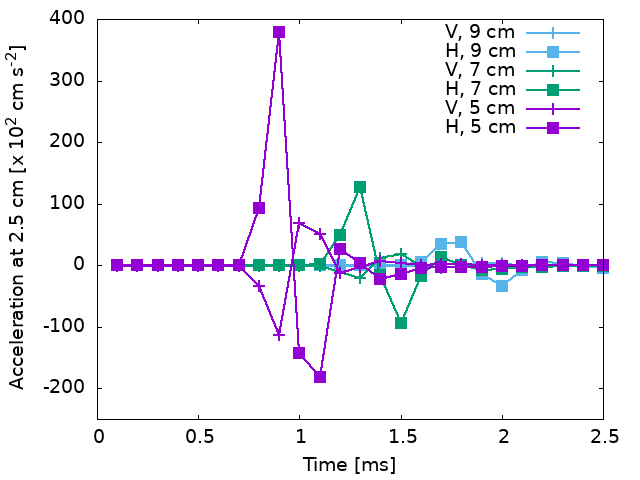}
    \caption{Acceleration at depths of 0.5 cm (upper panel) and 2.5 cm (lower panel) with elapsed time for Run 5. Blue, green, and purple represent the measured positions at distances of 9, 7, and 5 cm from the impact point, respectively. Cross and square symbols represent vertical and horizontal acceleration, respectively.}
    \label{fig:wavechk}
\end{figure}

\begin{figure}
    \centering
    \includegraphics[bb=0 0 536 378, width=10cm]{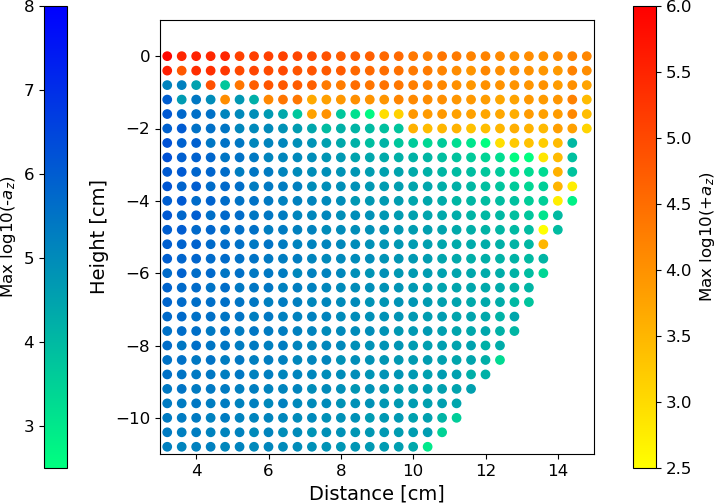}
    \caption{Maximum vertical acceleration distribution for Run 5. Color bars represent the logarithm of the maximum vertical acceleration ($a_z$). The warm-color plots represent acceleration in the upward direction, whereas the cold-color plots represent acceleration in the downward direction.}
    \label{fig:wavedistri}
\end{figure}

The results are presented for two measured accelerations at depths of 0.5 cm and 2.5 cm.
Figure~\ref{fig:wavechk} shows the time-lapse of the vertical and radial components of the acceleration. The sampling points were 5, 7, and 9 cm from the impact point.
The acceleration directions of the vertical components differed owing to shock propagation, where the shallower region was directed upward, and the deeper region was directed downward.
As shown in Figure~\ref{fig:pandvec}, the direction of the velocity was upward near the surface zone, but was downward at the deeper region.
Figure~\ref{fig:wavedistri} shows the vertical acceleration distribution.
The direction of the maximum vertical acceleration changed near the surface owing to the free surface.
Figure~\ref{fig:wavedistri} shows the maximum vertical acceleration distribution for Run 5. The phase change indicates the boundary of the near-surface zone.
Similar trends were observed in the experiments.
\citet{Matsue2020} set the accelerometer near the surface and detected the upward acceleration, whereas \citet{Quillen2022} set the accelerometer at the bottom of the container and detected the downward acceleration.
The present simulation clearly shows a difference in the directions. 

\begin{figure}
    \centering
    \includegraphics[bb=0 0 480 360, width=10cm]{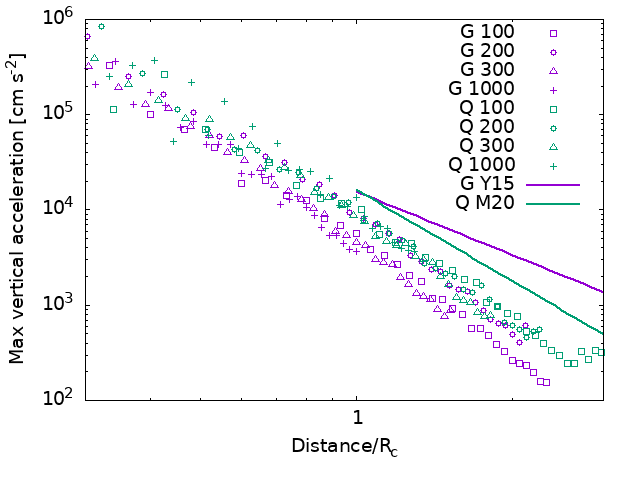}
    \caption{Maximum vertical acceleration at a depth of 0.5 cm versus the distance normalized by the crater radius. Purple and green points represent simulation results for glass and quartz, respectively. Square, circle, triangle, and cross symbols represent impact velocities of 100, 200, 300, and 1000 m s$^{-1}$, respectively. Purple and green lines are experimental results fitted for glass \citep{Yasui2015} and quartz \citep{Matsue2020}, respectively.}
    \label{fig:shake}
\end{figure}

Figure~\ref{fig:shake} shows the impact-induced acceleration for the simulation at a depth of 0.5 cm.
The maximum values of the vertical acceleration corresponded to the measured values of the impact-induced seismic wave. 
Growth of the crater stopped at a maximum vertical acceleration of $\sim10^{4}$ cm s$^{-2}$, corresponding to ten times the gravitational acceleration.
A similar situation was also confirmed in the experiments, although the peak acceleration from the simulation was smaller than the experimental results by a factor of two.
The simulated peak acceleration is significantly lower than the experimental values. This discrepancy is attributed to the pressure response to the density variation in the granular equation of state.
Approximating the equation of state as $P(\rho) = P_0 \rho^\gamma$, the acceleration scales as $a = - \nabla P(\rho) /\rho \sim -\gamma P/\rho^2 \nabla \rho$, indicating that the exponent $\gamma$ governs the magnitude of acceleration.
Because the developed model does not account for grain crushing or changes in the coordination number, the effective $\gamma$ is likely underestimated.
Because analysis of the impact-induced seismology can facilitate direct detection of the shock propagation determined by the properties and the interior structure of a material, the model can be used to predict the surface features of asteroids in future exploration.

In the simulations, the impact-induced oscillations for the glass-bead target were almost the same for the quartz sand target.
In contrast, the experimental results showed that the oscillation for glass was less damped than that of quartz.
This difference is owing to the granular target model used in the present simulations, where the sound velocity of glass is slightly different from that of quartz.
The Poisson’s ratio of the granular target used in the present simulation is 0.25, whereas that of real granular targets is approximately 0.3, which means that the physical properties of the granular targets differed from those of real granular targets.
Further study is needed to construct a suitable model of the granular target.

\begin{figure}
    \centering
    \includegraphics[bb=0 0 480 360, width=10cm]{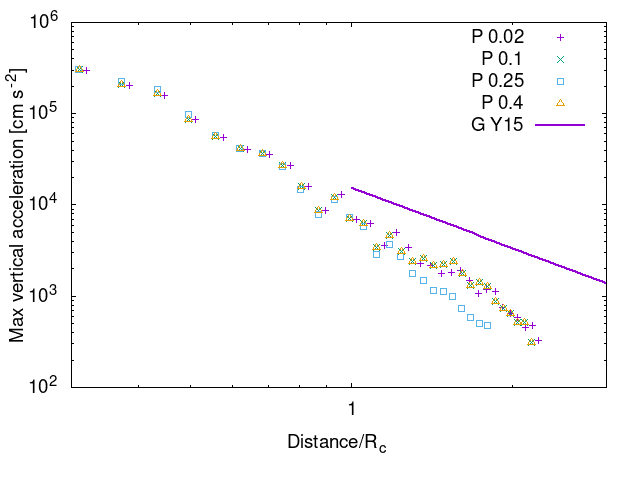}
    \includegraphics[bb=0 0 480 360, width=10cm]{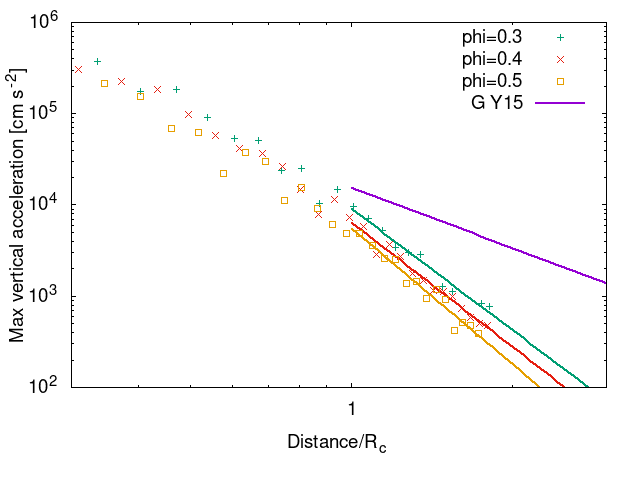}
    \caption{Same as Figure~\ref{fig:shake} but with different Poisson’s ratios (upper panel) and initial porosities (lower panel). The granular target is composed of soda-lime glass but the Poisson’s ratio or initial porosity is different. The impact velocity is 100 m s$^{-1}$. \\
    Upper panel: Purple, green, blue, and orange points represent the Poisson’s ratio of the target with values of 0.027, 0.1, 0.25, and 0.4, respectively. \\
    Lower panel: Green, red, and orange points represent the initial porosity of the target with values of 0.3, 0.4, and 0.5, respectively. Green, red and orange lines are fitting results for the numerical simulations corresponding to targets with an initial porosity of 0.3, 0.4, and 0.5, respectively.}
    \label{fig:wavechk_param}
\end{figure}

As shown in Table~\ref{tab:porpoi}, the radius of the crater was also affected by the Poisson’s ratio and the initial porosity of the granular target.
Thus, the Poisson’s ratio and the initial porosity also affect the impact-induced seismic wave.
Figure~\ref{fig:wavechk_param} shows the peak acceleration for different Poisson’s ratios and initial porosities for the model employing soda-lime glass.
The results indicate that increasing the Poisson’s ratio leads to slightly reduced damping, suggesting that inter-particle friction, implicitly represented by Poisson’s ratio, affects the energy dissipation.
Additionally, the initial porosity had a measurable influence on the damping behavior. Specifically, the slope of the peak acceleration attenuation with distance was $-4.42$, $-4.52$, and $-4.94$ at a porosity of 0.1, 0.4, and 0.5, respectively.
These results demonstrate that higher porosity leads to stronger acceleration damping.

\subsection{Analytical estimate of the crater}

Impact cratering is accompanied by impact-induced shaking, which is an important consideration for the crater region.
This subsection discusses the analytical estimate of the region where the impact-induced pressure exceeds the yielding strength of the granular material.
The condition $P(R) \ge Y$ defines the region where the impact-induced shock accelerates the granular material, but not all of this material escapes the target.
The impact wave should stop when the condition $P(R) = Y$ is satisfied, where $P(R)$ is the pressure at a distance from the impact point $R$ and $Y$ is the material strength of the soil target.
The pressure at the impact point should be determined by the isobaric core \citep[e.g.,][]{Miyayama2024}.
We assumed the pressure attenuation as:
\begin{equation}
    P =  P_{c,0} \left( \frac{v}{c_{s,0}} \right)^\beta \left( \frac{l}{r_p} \right)^{-\alpha} \label{shock}
\end{equation}
where $P_{c,0}$ is the pressure at the isobaric core when the impact velocity is $c_{s,0}$, $v$ is the impact velocity, $r_p$ is the radius of the projectile, $c_{s,0}$ is the speed of sound when the pressure is $P_0 = \rho_0 g r_p$, $l$ is the distance from the impact point, $\alpha$ is the shock attenuation parameter, and $\beta$ is the power exponent for the impact velocity.
The material strength of the soil is defined as :
\begin{equation}
    Y = f \rho_0 g h + f P \label{soil}
\end{equation}
where $g$ is the gravity, $h$ is the depth from the surface, and $f$ is the friction coefficient.
By combining Eqs.~\ref{shock} and \ref{soil} and applying the condition $P=Y$, we obtain the following expression:
\begin{equation}
    \left(\frac{l}{r_p}\right)^{\alpha} = \left( \frac{1}{f}-1 \right) \frac{P_{c,0}}{\rho_0 g h}  \left( \frac{v}{c_{s,0}} \right)^{\beta}. \label{ldcrater}
\end{equation}
We assume the seismic wave region as the zone where the depth is equal to the projectile radius $r_p$, i.e., $h = r_p$. 
The assumption $l = R_w$ corresponds to the typical radial distance at which the impact-induced pressure falls below the yield strength of the granular material. These definitions are used to estimate the spatial extent of the wave-affected region in our analytical model.
Thus, we found an analytical value for the seismic wave region $R_w$ by substituting $l=R_w$ and $h=r_p$ into Eq.~\ref{ldcrater}:
\begin{equation}
    \frac{R_w}{r_p} =  \left[ \left( \frac{1}{f}-1 \right) \frac{P_{c,0}}{\rho_0 g r_p} \right]^{\frac{1}{\alpha}} \left( \frac{v}{c_{s,0}} \right)^{\frac{\beta}{\alpha}}. \label{anal-crater}
\end{equation}
Eq.~\ref{anal-crater} is represented by $\pi$-scaling equations as follows:
\begin{eqnarray}
     \frac{\pi_R}{\pi_4^{b}}& = &  \frac{r_p}{\pi_4^b} \left(\frac{\rho_t}{m_p}\right)^{1/3} \left[ \left( \frac{1}{f}-1 \right) \frac{P_{c,0}}{\rho_0 g r_p} \right]^{\frac{1}{\alpha}} \left( \frac{g r_p}{c_{s,0}^2} \right)^{\frac{\beta}{2\alpha}} \left( \frac{gr_p}{v^2} \right)^{-\frac{\beta}{2\alpha}} \label{piscaleanal} \\
    \pi_R &=&  R \left(\frac{\rho_t}{m_p}\right)^{1/3} \\
    \pi_4^{b} &= & \left( \frac{\rho_t}{\rho_p}  \right)^{\frac{\beta}{6\alpha}}
\end{eqnarray}
The impact-induced wave region does not coincide with the ejection region.
Although the shock generates a velocity field, the kinetic energy of particles located beyond the excavation region is insufficient to overcome gravity and material cohesion.
Thus, these particles remain within the target.
From the analytical expression in Eq.~\ref{anal-crater}, the estimated radial extent of the shock-affected region $R_w$ is approximately ten times larger than the crater radius $R$ predicted by the same $\pi_2$ scaling.
For the typical parameter values used in our simulations, this ratio is approximately 10.
This reflects the fact that perturbation-induced granular fluidization extends far beyond the region of material excavation.

\begin{figure}
    \centering
    \includegraphics[bb=0 0 480 360, width=10cm]{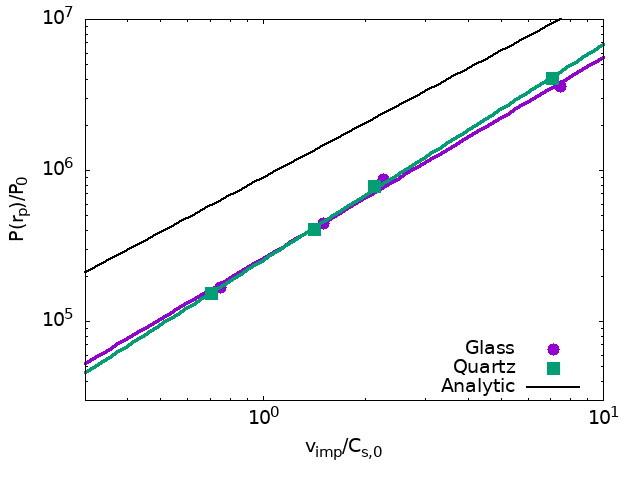}
    \caption{Relationship between isobaric core pressure and impact velocity. Purple and green points are glass and quartz results, respectively. Lines are fitting results using $P(r_p)/P_0=A (v_\mathrm{imp}/c_{s,0})^{a}$. The purple line corresponds to glass, where $A=10^{5.41}$ and $a=1.33$. The green line corresponds to quartz, where $A=10^{5.41}$ and $a=1.42$. The black line shows the analytical solution of Eq.~\ref{Pc-form}.}
    \label{fig:pc-anal}
\end{figure}

\begin{figure}
    \centering
    \includegraphics[bb=0 0 480 360, width=10cm]{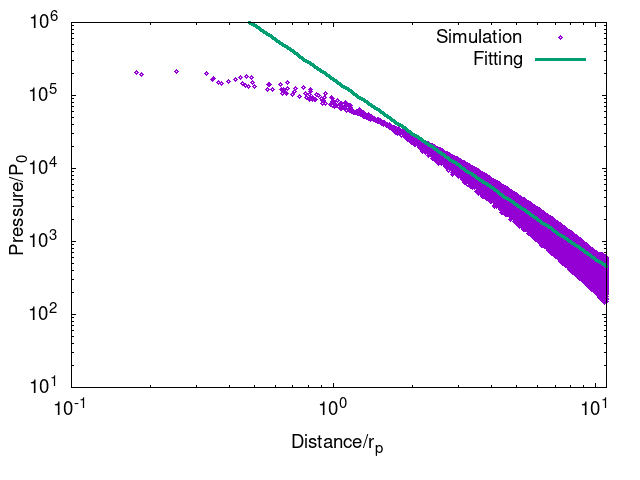}
    \includegraphics[bb=0 0 480 360, width=10cm]{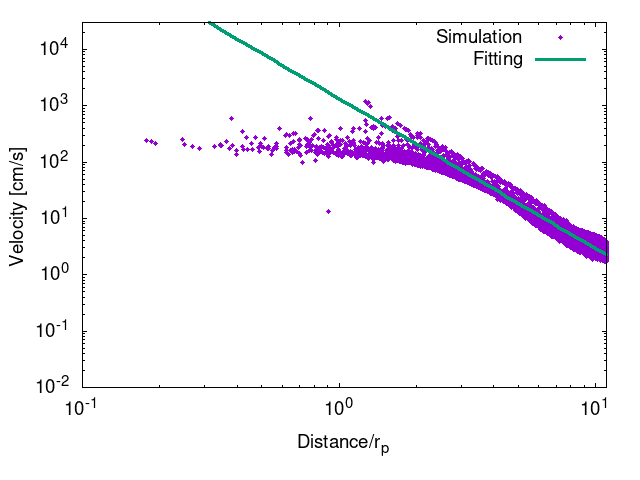}
    \caption{Upper panel: Relationship between the pressure and distance from the center of the impact point normalized by the projectile radius for Run 1 at 3 ms. Purple points show the pressure for SPH particles. The green line shows the fitting results for $P/P_0 = 10^{5.22} (l/r_p)^{-2.46}$. Lower panel: Relationship between velocity and distance. The green line shows the fitting results for $v=10^{2.98} (l/r_p)^{-2.51}$ cm s$^{-1}$.}
    \label{fig:presvec}
\end{figure}

Eq.~\ref{ldcrater} was derived under the assumption that the pressure distribution generated by the impact can be expressed as a power law function of the impact velocity and radial distance, as formulated in Eq.~\ref{shock}.
To validate this assumption, we extracted the velocity dependence and spatial attenuation of the pressure from the simulation results.
First, we examined the velocity dependence of the peak pressure at the impact point, corresponding to the isobaric core.
Figure~\ref{fig:pc-anal} shows the relationship between the pressure and impact velocity at the isobaric core.
By fitting the data to the form $P(r_p)/P_0 = A (v/c_{s,0})^\beta$, we obtained $\beta = 1.33$ for glass and $\beta = 1.42$ for quartz, indicating that the pressure scales well with the impact velocity according to a power law.
Thereafter, we analyzed the radial attenuation of the pressure and velocity.
Figure~\ref{fig:presvec} shows an example of the pressure and velocity profiles at the moment the shock front reaches the target boundary for Run 1.
The data were well approximated by fitting to power laws: $P/P_0 = 10^{5.22} (l/r_p)^{-2.46}$ and $v = 10^{2.98} (l/r_p)^{-2.51}$ [cm s$^{-1}$].
We extended this analysis to other parameter sets, fitting the pressure attenuation using the form $P/P_0 = 10^B (l/r_p)^\alpha$.
The results are summarized in Table~\ref{tab:sims}.
From the to our simulations, the mean values were $\alpha = -2.50$, $\beta = 1.33$ for glass, and $\alpha = -2.53$, $\beta = 1.42$ for quartz.
These attenuation exponents are slightly larger than -3, which corresponds to the constant late-stage equivalent energy condition \citep[e.g.,][]{Mizutani1983,Mizutani1990}.

We examined the same dependencies by developing an analytical model based on Eq.~\ref{shock}.
The initial peak pressure is approximated by $P_{c,0} \approx \frac{1}{2} \rho_0 c_s v$, assuming that the impact velocity is comparable to the speed of sound in the granular target. 
According to Eq.~\ref{Keff1}, the sound speed is pressure-dependent and follows $c_s \propto P^\varphi$, where $\varphi = 1/6$.
Substituting this relation into Eq.~\ref{shock} yields:
\begin{equation}
\frac{P_{c,0}}{P_0} = \left[\frac{\rho_0 c_{s,0}^2}{2P_0}\right]^{\frac{1}{1-\varphi}} \left( \frac{v}{c_{s,0}} \right)^{\frac{1}{1-\varphi}}, \label{PI-eq}
\end{equation}
where $P_0 = \rho_0 g r_p$ is the reference pressure and $c_{s,0}$ is the speed of sound at $P_0$.
According to theoretical analysis, the pressure attenuation with distance follows a power-law decay with an exponent $\alpha = 3$ \citep[e.g.,][]{Kieffer1980, Mizutani1990}.
Incorporating this attenuation into Eq.~\ref{PI-eq} gives:
\begin{equation}
\frac{P_{c,0}}{P_0} = \left( \frac{\rho_0 c_{s,0}^2}{2P_0} \right)^{6/5} \left( \frac{v}{c_{s,0}} \right)^{6/5} \left( \frac{l}{r_p} \right)^{-3}. \label{Pc-form}    
\end{equation}
This analytical model corresponds to a velocity exponent $\beta = 6/5$ and a spatial attenuation exponent $\alpha = 3$. 
We clarified the distinction between these two parameters: $\beta$ describes the velocity dependence of the peak pressure, whereas $\alpha$ characterizes the radial decay of the pressure.
Eq.~\ref{Pc-form} is presented prior to the introduction of Figure~\ref{fig:pc-anal}, which provides a comparison of the results from the analytical model with those from the simulation.
As shown in Figure~\ref{fig:pc-anal}, the analytical model overestimates the peak pressure by a factor of approximately 2, and the slope is slightly shallower than that observed in the simulations.
This discrepancy likely arises because the analytical model does not fully account for energy dissipation near the impact point.

\subsection{Dependence on target size and resolution} \label{app1}
\begin{figure}
    \centering
    \includegraphics[bb=0 0 480 360, width=\linewidth]{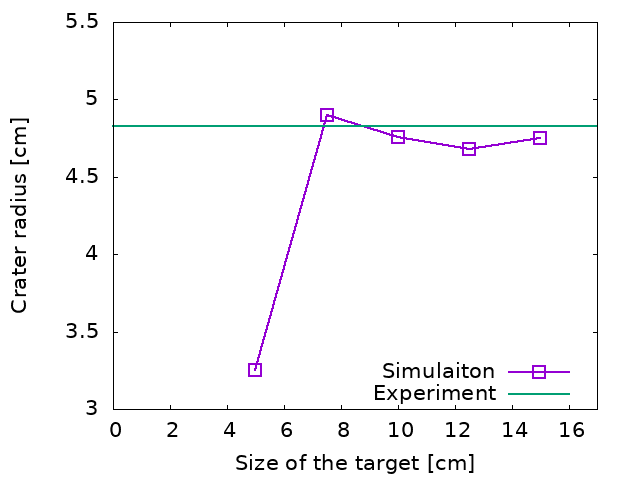}
    \caption{Relationship between the size of the target and the crater radius. Purple squares indicate SPH simulations. Green line shows the crater size estimated by \citet{Matsue2020}.}
    \label{fig:sizechk}
\end{figure}

\begin{figure}
    \centering
    \includegraphics[bb=0 0 480 360, width=\linewidth]{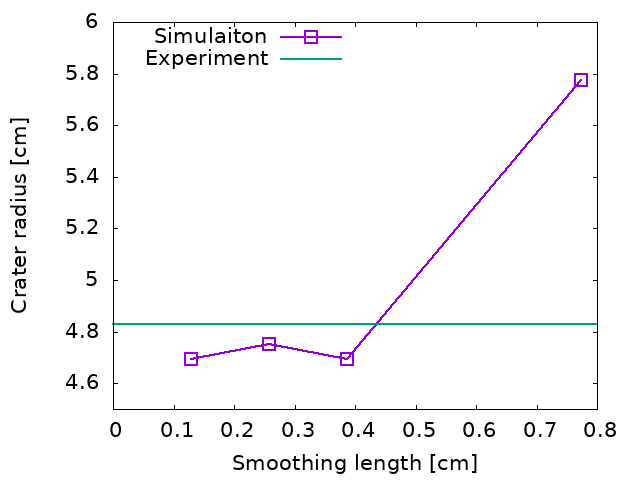}
    \caption{Relationship between smoothing length and crater radius. The colors are identical to those in Figure~\ref{fig:sizechk}.}
    \label{fig:resolutionchk}
\end{figure}

This subsection summarizes the tests used to validate the target size and resolution using the quartz target as an example.
The impactor is composed of a polycarbonate cylinder with a height and radius of 1 and 0.5 cm, respectively, with an impact velocity of $100$ m s$^{-1}$. 
In this case, the rim-to-rim crater radius is 4.8 cm, as predicted by \citet{Matsue2020}.
The numerical results are summarized in Table~\ref{tab:ressize}.
Figure \ref{fig:sizechk} shows the effect of the size of the hemisphere-shaped target.
We compared targets for which the radius of the hemisphere is 0.5, 7.5, 10, 12.5, and 15 cm, respectively. 
All targets had a smoothing length of 0.258 cm.
The crater size was not dependent on the boundary when the target radius was 1.5 times larger than the resultant crater size.
When the target size was comparable to the predicted crater size, the resultant crater size decreased owing to the boundary. 
Figure~\ref{fig:resolutionchk} shows the dependence of the numerical resolution.
A smaller smoothing length indicates a higher numerical resolution. 
The crater sizes from the simulation converged when the smoothing length was smaller than 0.4 cm, corresponding to the impactor radius.
These results highlight two conditions for reproducing crater formation by the SPH.
The target size should be 1.5 times larger than the resultant crater size, and the smoothing length should be smaller than the impactor radius.

For subsonic impact conditions, the crater size was found to converge sufficiently with increasing resolution, indicating that the present model is effective for analyzing low-velocity impacts.
In contrast, supersonic impacts are expected to produce larger craters and involve regions where grain crushing may occur. Therefore, higher-resolution simulations and larger target domains would be necessary to accurately capture the associated pressure fields and material model under such conditions.

\begin{table}[htbp]
    \centering
    \begin{tabular}{ccccccc}
    \hline
    Run & Size [cm] & $h$ [cm] & $R_\mathrm{rim}$ [cm]  \\
    \hline
    9 & 5.0  & 0.258 & 3.25  \\
    10& 7.5  &       & 4.90  \\
    11& 10.0 &       & 4.76   \\
    12& 12.5 &       & 4.68  \\
    13& 15.0 &       & 4.75  \\
    \hline                         
    14& 15.0 & 0.773 & 5.78  \\
    15&      & 0.387 & 4.69  \\
    16&      & 0.129 & 4.69  \\
    \hline
    \end{tabular}
    \caption{Simulation results for impacts on quartz sand at a velocity of $100$ m s$^{-1}$ for various target sizes and resolutions. Size corresponds to the target size, and $h$ is the smoothing length. For Runs 9--13, the smoothing length was 0.258 cm. For Runs 14--16, the target size was 15.0 cm.}
    \label{tab:ressize}
\end{table}

\section{Conclusion} \label{sec:conclusion}
Crater formation and impact-induced seismic waves on a granular material were studied using numerical simulations.
We constructed a sand target based on a close-packed granular model and carried out impact simulations using Smoothed Particle Hydrodynamics with a material strength parameter to understand the process of formation of impact craters on sand targets.
The sand target was assumed to be a granular material composed of soda-lime glass or quartz spheres, and the effective bulk and shear modulus of the granular material were derived.
The sound velocities of the glass and quartz sand were investigated and found to be consistent with those estimated from experiments.
Thus, the granular target model is adequate for the simulations.
The crater radius and impact-induced seismic wave on the glass or quartz target were investigated using the SPH with the material strength parameter. 
We compared the simulation results with experimental data based on the $\pi$-scaling law and found that the powers of $\pi_2$ determined by the simulations (glass: $-0.189$; quartz: $-0.179$) were consistent with the experimentally determined values (glass: $-0.19$; quartz:$-0.18$).
We analyzed the effective soil strength and peak pressure at the initial contact and compression stage to understand the mechanism related to the power of $\pi_2$ and found that the ratio representing the velocity dependence of the peak pressure and the shock attenuation controlled the power of the $\pi$-scaling in the gravity regime.
We also investigated the maximum acceleration during crater formation and compared the data with the experimental values determined by measuring the impact-induced seismic wave.
The maximum acceleration determined by the simulations agreed with that measured for the impact-induced seismic wave, although the numerical results were a factor of two smaller than the experimental values.
The simulations indicate that crater growth stopped at the distance where the acceleration was smaller than 10 times the gravitational acceleration.
This phenomenon was also confirmed by the experimental results.
We also found that the direction of vertical acceleration varied with depth, owing to differences in the shock propagation at different measurement depths.
We successfully constructed a model for future exploration and experimental evaluation of the impact on asteroids by comparing numerical simulations and experiments using a cohesionless regolith material.
The model will be useful for investigating the surface properties of asteroids in the analysis of exploration.


\section*{Acknowledgments}
The authors thank K. Sugiura for developing and making publicly available the SPH simulation code \citep{Sugiura2018} used in this study.
Numerical computations were carried out on Cray XC50 and XD2000 at the Center for Computational Astrophysics, National Astronomical Observatory of Japan, and computational resources provided by National Defense Academy. This work is supported by JSPS Grants-in-Aid for Scientific Research No. 24K07114 (K.K.), 22H00179 (M.A.), 23H01231 (K.K.).




\end{document}